\documentclass[pre,twocolumn,showpacs]{revtex4}
\usepackage{graphicx}
\usepackage{amsmath}
\begin{document}
%
\title{Epidemic spreading with immunization and mutations}
\author{Stephan~M.~Dammer$^{1,2,}$\footnote[1]{Corresponding author: {\tt
      dammer@comphys.uni-duisburg.de} .} and Haye Hinrichsen$^{2}$}
\affiliation{$^{1}$ Institut f\"ur Physik,
        Universit{\"a}t Duisburg-Essen, 47048 Duisburg, Germany}
\affiliation{$^{2}$ Theoretische Physik, Fachbereich 8,
        Bergische Universit{\"a}t Wuppertal, 42097 Wuppertal, Germany}
\begin{abstract}
The spreading of infectious diseases with and without immunization of
individuals can be modeled by stochastic processes that exhibit a
transition between an active phase of epidemic spreading
and an absorbing phase, where the disease dies out. 
In nature, however, the transmitted pathogen may also mutate, 
weakening the effect of immunization. In order to study the influence of
mutations, we introduce a model that mimics epidemic spreading 
with immunization and mutations. The model exhibits a line of
continuous phase transitions and includes the general epidemic process (GEP) 
and directed percolation (DP) as special cases. Restricting 
to perfect immunization in two spatial dimensions we analyze the phase diagram and study
the scaling behavior along the phase transition line as well as in the vicinity
of the GEP point. We show that mutations lead generically to a crossover 
from the GEP to DP. Using standard scaling arguments we also predict the 
form of the phase transition line close to the GEP point.  
It turns out that the protection gained by immunization is
vitally decreased by the occurrence of mutations.
\\
\end{abstract}
\maketitle
\def\xvec{\text{\bf x}}
\parskip 1mm 
%
\section{INTRODUCTION}
\label{intro}

The modeling of epidemic spreading is a fascinating subject 
both in theoretical biology and statistical physics far from 
equilibrium~\cite{Daley,Mollison77}. A possible approach 
is the study of stochastic models that mimic the competition of
infectious spreading and recovery by certain probabilistic rules. 
Depending on the rates for infection and recovery the disease may either
spread over the population or disappear after some time. 

The simplest models for epidemic spreading 
assume that the individuals live on 
the sites of a $d$-dimensional lattice. 
At a given time each individual can be either infected or healthy.
The system evolves according to certain probabilistic rules that resemble
infection of nearest neighbors and spontaneous recovery. If the susceptibility 
to infections is sufficiently large the epidemic will spread while for low
susceptibilities it will die out.

In most models it is assumed that the contagious disease is transmitted 
exclusively by direct contact. This means that the disease, 
once extinct, cannot appear again. 
The system is then trapped in a fully recovered state, which
can be reached but not be left. Such a state, where the system 
is dynamically trapped, is called {\it absorbing}.
The two regimes of spreading and extinction are usually
separated by a so-called absorbing phase transition. 
In the supercritical regime, where infections dominate,
the epidemic may spread over the entire system, reaching a 
fluctuating stationary state with a certain average density of infected
individuals. In the subcritical regime, where recovery dominates, the system
eventually reaches the fully recovered absorbing state.

Close to the transition the temporal evolution of the spreading process is
characterized by large-scale fluctuations. Theoretical interest
in epidemic spreading stems from the fact that this type of critical behavior
is universal, i.e., it does not depend on the details of the model under consideration.
The classification of all possible transitions from fluctuating phases
into absorbing states is currently one of the major goals of nonequilibrium statistical
physics~\cite{MarroDickman,Hinrichsen}. 

In epidemic models without immunization, where the pathogen (e.g.~a virus) is transmitted
to nearest neighbors, the critical behavior at the transition generically
belongs to the universality class of directed percolation (DP)~\cite{Kinzel85}. 
The DP class is very robust and plays an important role not 
only in epidemiology but also in various other
fields such as Reggeon field theory~\cite{Reggeon}, 
interface depinning~\cite{TakayasuTretyakov92}, population growth~\cite{Grassberger83},
catalytic reactions~\cite{ZGB86}, and flowing sand~\cite{HJRD99} (for a 
review of possible experimental applications see~\cite{HinrichsenBrazil}).

As a next step towards a more realistic description of epidemic spreading one can
take the effect of immunization or weakening by infections
into account~\cite{Grassberger83,CardyGrassberger85,Janssen85,GCR97,Andrea}. The simplest
way of implementing immunization is to change the initial susceptibility 
of an individual after the first infection and keep 
it constant thereafter. Generally such a process
is controlled by two parameters, namely, a probability for first infections 
and a reinfection probability. 
In the case of perfect immunization, where each individual can be 
infected only once, one obtains the so-called `general epidemic process' (GEP)~\cite{Daley,Mollison77,Grassberger83}. It differs from DP in so
far as the disease can only spread in those parts of the system which have not
been infected before. Thus, starting with a single infected site in a
non-immune environment, the  disease typically propagates in 
form of a solitary wave, leaving a region 
of immune sites behind. Depending on the infection rate this wave may
either spread over the entire system or stop before. The transition between 
infinite and finite spreading is a critical phenomenon which in this case
belongs to the universality class of dynamical isotropic percolation~\cite{Stauffer}. 
Note that unlike DP models, a GEP running on a {\em finite} system has no fluctuating
active state, instead the process terminates when it reaches the boundaries.

In nature, however, immunization is a much more complex
phenomenon. For example, the protection by immunization may abate
as time proceeds. Even more importantly, the strategy of immunization 
competes with the ability of the contagious pathogen to mutate so that 
it can no longer be recognized by the immune system of previously infected
individuals, weakening the effect of immunization.
The aim of the present work is to introduce and study a simple
model which mimics epidemic spreading with immunization and
mutation. To this end we generalize the model described in~\cite{GCR97,Andrea} 
by including mutations as well as a mechanism for the competition between 
different species of pathogens. The model is controlled by
three parameters, namely, a first infection probability, 
a reinfection probability controlling the effect of immunization, and 
a probability for spontaneous mutations. 
Moreover, the model is defined in such a way 
that it includes DP and GEP as special cases.
    
The article is structured as follows. In Sec.~\ref{model} we first define 
the model. Restricting our analysis to the case of perfect 
immunization in two spatial dimensions, we discuss the phase diagram and the qualitative
behavior of the model in Sec.~\ref{phasediagram}. In Sec.~\ref{line} the
critical behavior at the phase transition line is studied in detail
while crossover phenomena in the vicinity of the GEP point are investigated
in Sec.~\ref{nearGEP}. The article ends with concluding remarks 
in Sec.~\ref{conclusion}.   
%
%
\section{SIMULATION MODEL}
\label{model}
Our model is meant to describe the spreading of an 
infectious disease that evolves as follows. Individuals can 
be healthy or infected with a certain pathogen. During their illness
infected individuals may infect or reinfect neighboring individuals 
with certain probabilities. Moreover, there is a probability that a 
pathogen mutates during transmission. 
Because of the enormous number of possible mutations
one usually obtains an entirely new type of pathogen which has not been 
involved before. To simplify the model we also assume that each
individual can be infected at a given time by no more than
a {\em single} type of pathogen. If the individual is exposed
simultaneously to several competing pathogens, one 
of them is randomly selected.

In more technical terms the model is defined as follows. 
Individuals live on the sites of a $d$-dimensional simple cubic lattice.
Each site may be either healthy (inactive) or infected (active) 
by a pathogen of type $n$, where $n$ is a positive integer. 
Moreover, each site keeps track of all species of pathogens by which 
it has been infected in the past. Therefore, the state of a site
is characterized by a number $n$ together with a dynamically generated list
of all previous types of infections.

The model evolves in time by synchronous updates, i.e., in each 
time step the whole lattice is updated in parallel as follows:
\begin{enumerate}

\item
{\bf Spreading:} Each infected individual at time $t$ transmits its pathogen $n$ 
to its $2d$ nearest neighbors (target sites) at time $t+1$.

\item
{\bf Susceptibility:} 
The transmitted pathogen reaches the target site at time $t+1$
with probability $p$ if this site has never been infected before 
by this species, otherwise it reaches the target site
with probability~$q$.

\item
{\bf Competition:}
If a target site is exposed to several transmitted pathogens, 
one of them is randomly selected with equal weight.

\item
{\bf Mutation and Infection:} Before infecting the target site the selected pathogen mutates
with probability $\lambda$, replacing $n$ by a new integer number (drawn from
a global counter) which has not been used before.

\item
{\bf Immunization:}
In case of a first infection the type of pathogen is added to the 
list of species against which the site will be immune in the future.

\item
{\bf Recovery:}
All sites that have been active at time $t$ recover until time $t+1$ unless they are
not again infected during the update from $t$ to $t+1$. Hence the time
of illness is a single time step.
\end{enumerate}
%
%
\section{PHASE DIAGRAM}
\label{phasediagram}

\begin{figure}
\includegraphics[width=80mm]{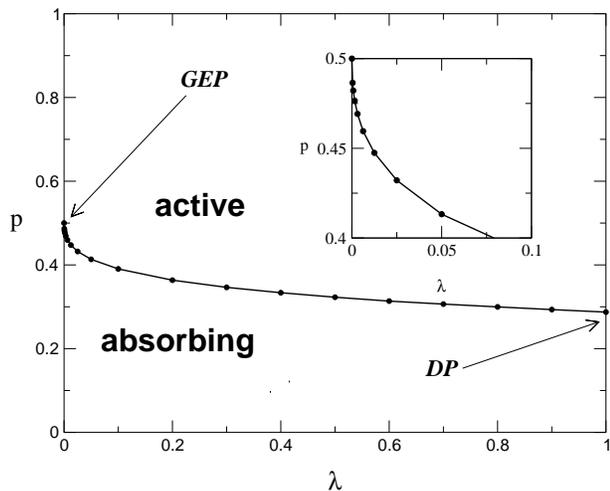}
\caption{
\label{figphasediagram}
Phase diagram of the model. The active and the inactive phases
are separated by a curved line of continuous phase transitions,
connecting the points of critical GEP and critical DP (see text). 
Circles mark the numerically determined critical points. 
The inset shows the phase transition line for small $\lambda$.
}
\end{figure}

In what follows we restrict our analysis to
the special case of perfect immunization $q=0$.
In this case the model is controlled by only two parameters, namely,
the probability of first infections $p$ and the probability of
mutations~$\lambda$. Moreover, we restrict ourselves to the case
of $d=2$ spatial dimensions. The corresponding phase diagram is shown in
Fig.~\ref{figphasediagram}. It comprises an active phase, where the
epidemic spreads, and an inactive phase, where the disease dies out
so that the system eventually enters the fully recovered absorbing state.
Both phases are separated by a curved phase transition line.

We first note that the endpoints of the phase transition line correspond
to well-known special cases. On the one hand, for $\lambda = 0$ 
the model reduces to the GEP on a square 
lattice~\cite{Grassberger83}, provided that only one type of 
pathogen is involved. In this case the critical value of $p$ is 
exactly given by $p_{\rm c} = 1/2$~\cite{Stauffer}. 
On the other hand, for $\lambda = 1$ all transmitted pathogens mutate, i.e.,
the target sites are always infected with a new species so that 
immunization has no influence. It is easy to see that in this 
case the model reduces to directed bond
percolation on a square lattice with 
$p_{\rm c} \simeq 0.28734$~\cite{Grassberger96}. The
other points on the phase transition line in Fig.~\ref{figphasediagram} 
were determined numerically using seed-simulations (see Sec.~\ref{seedsim}).

As already discussed in the Introduction, mutations generally weaken the
effect of immunization. This explains why the critical value of $p$ decreases
monotonically with increasing mutation probability $\lambda$.
Fig.~\ref{figphasediagram} clearly shows that introducing
mutations in a GEP has a strong influence on $p_{\rm c}$, especially
if $\lambda$ is very small. Since the phase transition line approaches 
the GEP point with infinite slope (see inset in Fig.~\ref{figphasediagram})
a tiny increase of $\lambda$ reduces the corresponding critical 
value of $p$ dramatically. This indicates that mutations are a relevant
perturbation and thus the spreading behavior of the model for $\lambda > 0$ 
is expected to differ from that of the GEP. On the other hand,
for larger values $\lambda\gtrsim 0.1$ the critical value 
of $p$ decreases only moderately with increasing $\lambda$, indicating
that the behavior of the model in this region is essentially the same
as for $\lambda = 1$, where the transition is known to belong to DP.      

Fig.~\ref{simpic} shows snapshots of simulations at the critical point
for different times and
various values of $\lambda$. Infected individuals are represented 
by black dots. Moreover, individuals which are immune against at least 
one active type of pathogen are marked by gray dots. If a site is immune
solely against pathogens that already became extinct the gray dot is
removed. The snapshots shown in this figure suggest the following
qualitative behavior:
\begin{figure}
\includegraphics[width=85mm]{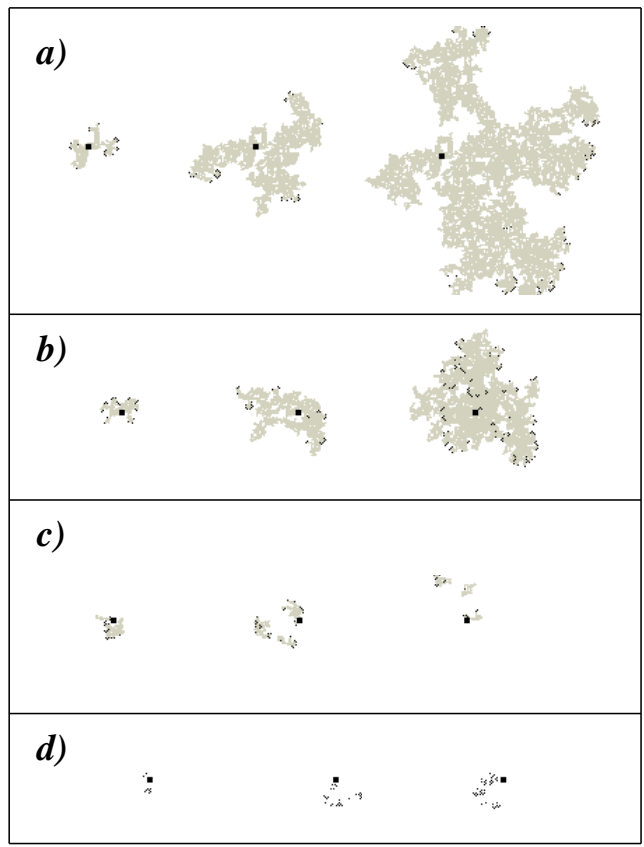}
\vspace*{-4mm}
\caption{
\label{simpic}
Snapshots of seed-simulations at the phase transition. The black square marks
the position of the seed. Each row shows a simulation
for a fixed value of $\lambda$ and times $t=30$, $t=100$ and $t=225$ (from
left to right). From $a)$ to $d)$ values of $\lambda$ are $0$ (GEP),
$0.003125$, $0.05$ and $1$ (DP). Black (gray) dots denote active (immune)
individuals (details in the text).  
}
\end{figure}
\begin{itemize}
\item[\bf a)]
For $\lambda = 0$ the spreading process
creates a growing cluster of immune sites with infected individuals
located at the edges. As usual in the GEP, the region of immune
sites is compact in the active phase while it is fractal at the phase transition. 
\item[\bf b)]
For a small value of $\lambda$ 
the disease first behaves like a
GEP (for $t\leq 100$) while for larger times the spreading behavior 
changes and clearly differs from the GEP. 
Although there still is a region of immune individuals it does
no longer provide efficient protection against infections since 
it is reinvaded by mutated pathogens. 
\item[\bf c)]
Increasing $\lambda$ further 
the process changes its appearance already at an early stage.
There are only small patches of immune sites and the process
reminds more of DP than GEP.
\item[\bf d)]
Finally for $\lambda = 1$ every transmitted pathogen mutates into a new one. 
In this case immunization has no influence and the process reduces to DP. 
Since the disease is not driven away from immunized
regions the spatial extension of an epidemic at the phase
transition grows much slower compared to a critical GEP. 
\end{itemize}
\noindent
Based on these phenomenological observations we expect the model to behave
initially in the same way as a GEP. After a certain time mutations
become relevant, allowing former immune areas to be reinvaded. 
Especially close to the transition the process survives long enough 
to reach this crossover time. The visual appearance of the process is then 
increasingly similar to that of a DP process. The time it takes 
to observe the crossover from GEP to DP grows with decreasing
$\lambda$ and eventually diverges in the limit $\lambda\rightarrow 0$.
%
%
\section{CRITICAL BEHAVIOR ALONG THE PHASE TRANSITION LINE}
\label{line}

Phase transitions into absorbing states are usually characterized by simple
scaling laws. In models for epidemic spreading an important quantity is
the probability $P_{\rm s}(t)$ that an epidemic starting from a single
infected seed in a healthy and non-immune environment survives at least 
until time~$t$.

Let us first recall the standard scaling laws for the survival probability.
Let $\triangle _{\rm p}= p-p_{\rm c}$ denote the distance from criticality. If
$|\Delta_{\rm p}| \ll 1$ the survival probability first decays as a power
law $P_{\rm s}(t) \sim t^{-\delta}$ until a certain timescale
$\xi _{||}\sim|\Delta_{\rm p}|^{-\nu _{||}}$ is reached from where on
it either saturates at $P_{\rm s}(\infty)\sim {\Delta_{\rm p}}^{\delta\nu _{||}}$
for ${\Delta _{\rm p}}>0$ or decays exponentially for ${\Delta_{\rm p}}<0$,
reaching a spatial extension  $\xi _{\perp}\sim|\Delta_{\rm p}|^{-\nu _{\perp}}$.
Assuming scaling invariance this behavior can be described in terms
of a scaling form
\begin{equation}
\label{scaling}
P_{\rm s}(t)=t^{-\delta}\Phi ({\Delta_{\rm p}}t^{1/\nu _{||}}) \,,
\end{equation}
where $\Phi (\zeta)$ is a scaling function with the asymptotic behavior
\begin{equation}
\Phi(\zeta) \sim  
\begin{cases} 
const & \text{ for } \zeta\to 0 \\ 
\zeta ^{\delta\nu _{||}} & \text{ for } \zeta\to+\infty \\
0 & \text{ for } \zeta\to-\infty 
\end{cases}
\end{equation}
such that the time dependence drops out for $\zeta \to +\infty$.
The scaling function $\Phi (\zeta)$ and the triplet of critical exponents
$(\nu _{||},\nu _{\perp},\delta)$ are believed to characterize the universality
class of the phase transition under consideration. We note that this scaling
form is known to be valid both for GEP and DP, although with different sets
of critical exponents~\cite{munoz2}
\begin{equation}\label{exponents}
(\nu _{||},\nu _{\perp},\delta)\simeq 
\begin{cases}
$(1.506, 4/3, 0.092)$ & \text{ for GEP } \\
$(1.295, 0.734, 0.451)$ & \text{ for DP. }
\end{cases}
\end{equation}
For the density $\rho$ in DP a similar scaling form as in Eq.~(\ref{scaling})
is valid (with the same exponents as in Eq.~(\ref{exponents}) for $d=2$).   

We now analyze the critical behavior of the model,
assuming that the scaling form~(\ref{scaling}) is valid
everywhere in the vicinity of the 
phase transition line. Although the
qualitative discussion in Sec.~\ref{intro} 
suggests DP behavior for
$0 < \lambda \leq 1$ we note that this would be a
non-trivial result since the so-called DP-conjecture does not
apply in the present case. 
The DP-conjecture~\cite{DPconjecture} states that phase transitions
in  two-state systems with a reachable absorbing state and
short-range interactions belong to DP, provided that memory effects,
non-conventional symmetries, and quenched disorder 
are absent. Contrarily the present model has
many absorbing states and memorizes previous infections over a long time.

\subsection{Seed simulations}
\label{seedsim}
%
%
\begin{figure*}
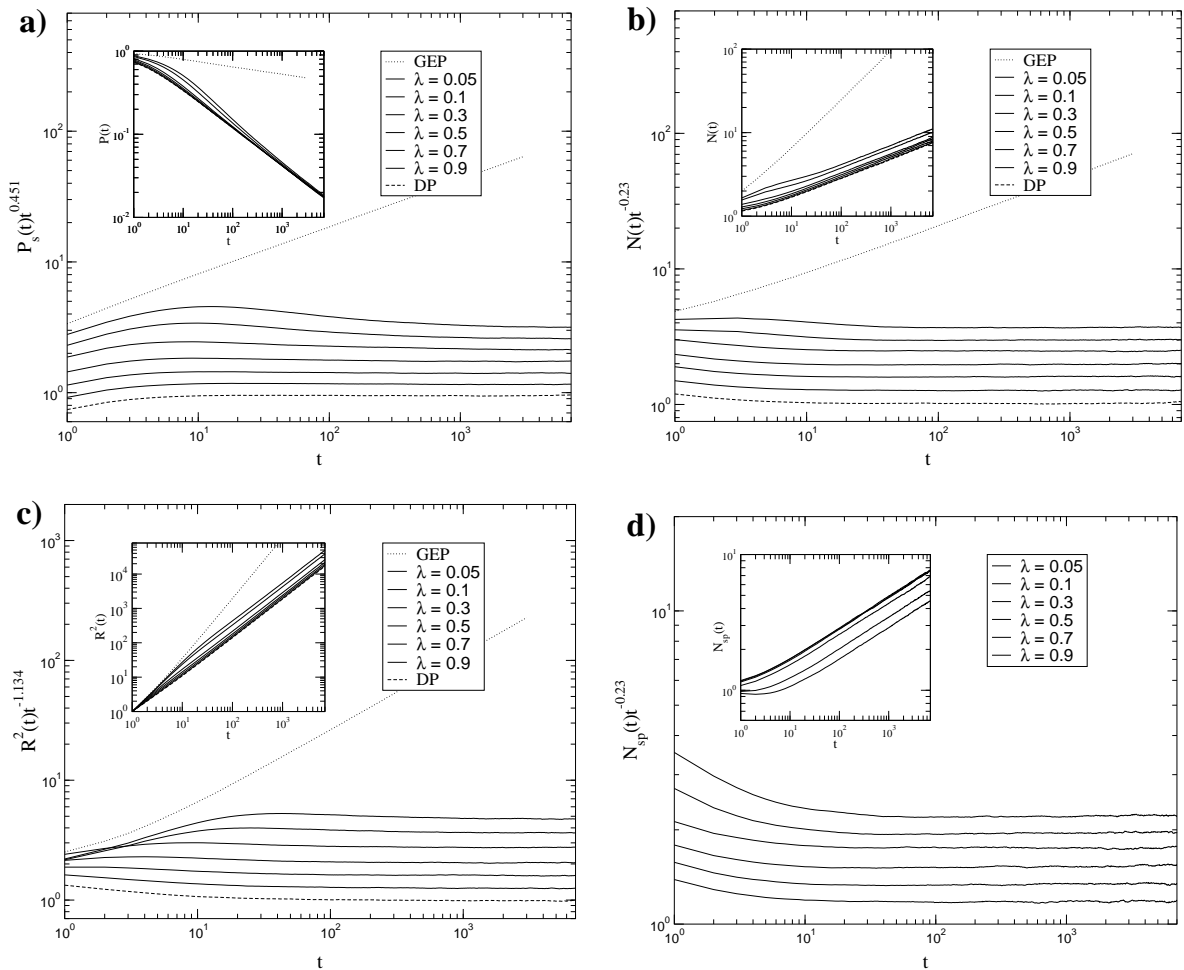
 
\includegraphics[width=75mm]{SURV.eps} \hglue 5mm
\includegraphics[width=75mm]{NA.eps}\vspace{4mm}\\
\includegraphics[width=75mm]{R_2.eps}\hglue 5mm
\includegraphics[width=75mm]{NSP.eps}
\caption{
\label{critseedpic}
Seed simulations:
Behavior of $P_{\rm s}(t)$, $N(t)$, $R^2(t)$ and $N_{\rm sp}(t)$ at the phase
transition for different values of $\lambda$. The data is multiplied
by the expected asymptotic power law and vertically shifted while the insets show the original data. 
}
\end{figure*} 

Seed simulations start with a single infected site at the origin 
in a  non-immune environment. Each run is stopped 
when it dies out or reaches a preset maximum
time. We average over many runs with different realizations of randomness. In order
to eliminate finite size effects the lattice is always chosen large enough so that 
the epidemic never reaches its boundaries. As usual in this type of simulation
we measure the survival probability $P_{\rm s}(t)$, the number of 
active sites  $N(t)$ averaged over all runs, and the mean square
spreading from the origin $R^2(t)$ averaged over all active sites in surviving
runs~\cite{GrassbergerTorre79}. The scaling form~(\ref{scaling}) implies that
these quantities vary at criticality algebraically as
\begin{equation}
\label{critseed}
P_{\rm s}(t)\sim t^{-\delta}\ \,,\ \ N(t)\sim t^\theta\ \ ,\ \ R^2(t)\sim t^{2/z}\ \,,
\end{equation} 
where $z=\nu_\parallel/\nu_\perp$ and $\theta=d/z-2\delta-1$ for GEP and
$\theta=d/z-2\delta$ for DP~\cite{munoz}.

Fig.~\ref{critseedpic} shows our results of seed-simulations along the phase
transition line. We note that these simulations are numerically challenging
since for each site a list of infections in the past has to be
created and continuously updated. Therefore, 
numerical simulations are mainly limited by the available memory
and restricted the present study to a maximal temporal range of about $10^4$ 
Monte Carlo updates.

In order to determine the critical threshold $p_{\rm c}(\lambda)$
we kept $\lambda$ fixed and varied $p$ until $P_{\rm s}(t)$
displayed the expected slope of DP in a log-log plot 
(dashed lines in Fig.~\ref{critseedpic}). 
Although this procedure is in favor of DP scaling the mere fact
that it works consistently for all quantities in Eq.~(\ref{critseed})
confirms that the transition does belong to DP for any $0<\lambda\leq 1$
while GEP scaling can be ruled out. The data also shows the
expected crossover. For small~$\lambda$ the curves roughly display 
the slope expected for GEP (dotted lines) before they cross over
to DP, confirming the crossover scenario discussed in the previous 
sections. Thus the introduction of mutations in a GEP is a relevant 
perturbation in the sense that it changes the asymptotic critical 
behavior of the model. 
At the phase transition the interplay 
between immunization and mutations drives the system towards DP.  

We note that the involvement of different species 
of pathogens in a spreading process with mutations
allows one to introduce the number of active species $N_{\rm sp}(t)$ as 
an additional order parameter. As shown in panel $d)$ of Fig.~\ref{critseedpic},
the number of active species at criticality increases in the same way as
the number of infected individuals. Thus their quotient tends to a $\lambda$-dependent 
constant
\begin{equation}
c(\lambda)=\lim_{t \to\infty} N/N_{\rm sp}\,.
\end{equation}

\subsection{Full-lattice simulations}
\label{fulllattice}

In this type of simulation we use a finite system of size 
$L\times L$ with periodic
boundary conditions. The initial configuration is a fully occupied 
lattice where all individuals are infected by different
types of pathogens. (It would be also possible to 
occupy all sites with the same type of pathogen, but then 
the process would immediately be trapped in the absorbing state.) 
We measure the density of active sites $\rho (t)$ and 
the density of active species $\rho _{\rm sp}(t)$. 
In the case of directed bond percolation $\lambda = 1$ it is known 
that the density of active sites decays as
$\rho (t)\sim t^{-\delta}$. Exemplarily we
performed a full-lattice simulation for
$\lambda = 0.5$ at the critical point, confirming this type of
decay with $\delta = 0.451$. Moreover, like in seed simulations, 
the density of active species also decays as $\rho
_{\rm sp}(t)\sim t^{-\delta}$. 

\begin{figure}
\includegraphics[width=80mm]{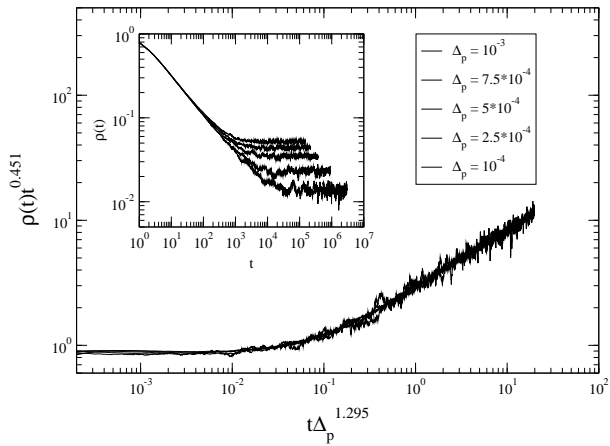}
\caption{
\label{figoffcrit}
Data collapse of $\rho (t)$ for $\lambda = 0.5$ using the critical
exponents of DP, i.e., $\delta = 0.451$ and $\nu_\parallel = 1.295$. 
}
\end{figure}

The three exponents $\delta,\theta,z$ in Eq.~(\ref{critseed}) that govern the
spreading behavior at the transition depend only on two of the three independent critical
exponents that are needed to characterize the universality class of DP (see
Eq.~(\ref{exponents})). In order to check the value of the third independent
exponent we performed off-critical full-lattice simulations for
$\lambda=0.5$ and different values of $0<\Delta_{\rm p}\ll 1$.  
As is shown in Fig.~\ref{figoffcrit} using the critical exponents $\delta$ and
$\nu_\parallel$ of DP one obtains a reasonable data collapse of the density $\rho
(t)$. This supports the claim that along the phase transition line for $\lambda > 0$
indeed all three exponents governing the scaling behavior of the spreading
process are that of DP.   
%
%
\section{CRITICAL BEHAVIOR IN THE VICINITY OF THE GEP POINT}
\label{nearGEP}
In this section we investigate the influence of mutations in the vicinity of
the GEP point in order to address two questions, namely, how does the system
cross over from GEP to DP and why does the phase transition line terminate
in the GEP point with an infinite slope.

\subsection{Decay at the GEP point}
\label{gepdecay}
\begin{figure}
\includegraphics[width=80mm]{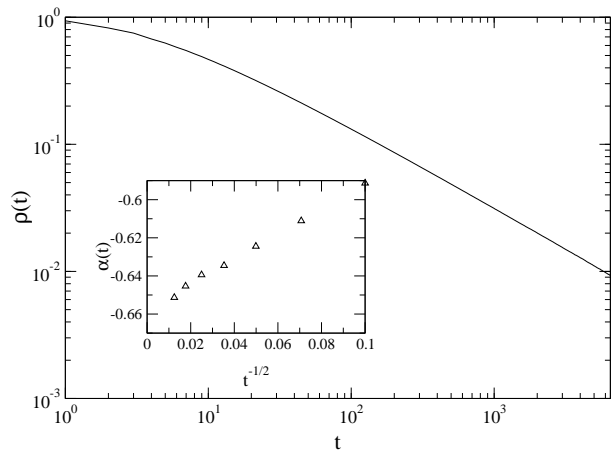} \vglue 4mm
\includegraphics[width=80mm]{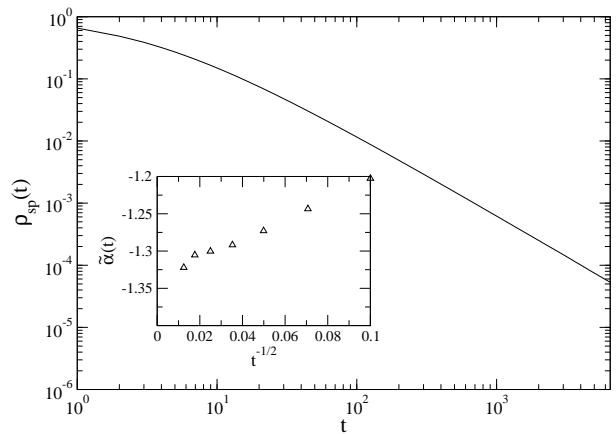}
\caption{
\label{figatgep}
Simulations at the GEP point $p=1/2, \lambda=0$ starting with a fully
occupied lattice, where each site carries a different
kind of infection. Although the lines are slightly curved,
the decay of $\rho$ and $\rho_{\rm sp}$ may suggest
a possible power-law behavior in the limit $t\to\infty$. As shown in the
insets, the corresponding exponents can be extrapolated visually by plotting
the effective slope against $1/\sqrt{t}$ (see text).
}
\end{figure}

Already at the GEP point our model exhibits a new feature,
namely, the competition of different types of infections
in full-lattice simulations (see Sec.~\ref{fulllattice}).
Starting the simulation in a finite system, where each site
is initially infected by a different kind of pathogen, one
observes a coarsening process of competing species, leading
to a slow decay of the density of active sites and active
species. As shown in Fig~\ref{figatgep}, the decay of these
quantities suggests possibly asymptotic power laws,
although the data display a considerable curvature in both
cases. {\em Assuming} asymptotic power laws 
\begin{equation}
\label{powerlawconjecture}
\rho(t) \sim t^{-\alpha}
\qquad
\rho_{\rm sp}(t) \sim t^{-\tilde{\alpha}}
\end{equation}
at the GEP point
we extrapolate the effective exponents $\alpha(t),\tilde{\alpha}(t)$ visually
for $t\to\infty$ (see insets of Fig.~\ref{figatgep}), obtaining
the estimates
\begin{equation}\label{alphas}
\alpha = 0.66(2)\,, \qquad \tilde{\alpha}= 1.33(4)\,,
\end{equation}
suggesting that $\tilde{\alpha} = 2\alpha$. 

Regarding the limited accuracy of our numerical simulations the 
conjecture of an asymptotic algebraic decay has to be taken with care.
However, indirect support comes from the one-dimensional case.
Here the GEP transition is shifted to $p_{\rm c}=1$ and the dynamics of
competing species reduces to a ballistic coalescence 
process~\cite{BallisticCoagulation}, for which asymptotic 
power laws could be derived exactly.

\subsection{$\lambda$-controlled transition at the GEP point}
\label{gepplusm}

Let us now turn to the critical behavior in the vicinity of the GEP point.
The GEP point in Fig.~\ref{figphasediagram} can be approached either
vertically by varying $p$ or horizontally by varying $\lambda$.
The critical behavior in vertical direction has been studied in
detail in Refs.~\cite{Grassberger83,CardyGrassberger85,Janssen85} and can be 
described in terms of the scaling form~(\ref{scaling}). Contrarily, moving
in horizontal direction by varying $\lambda$ and keeping $p=1/2$ fixed 
one encounters mutations as a new feature, leading to a non-trivial 
fluctuating active state.

As usual in critical phenomena, the additional control parameter $\lambda$
is associated with a novel critical exponent~$\mu_\parallel$.
Like $\nu_\parallel$ this exponent is defined in such a way
that
\begin{equation}\label{xi_par}
\xi_\parallel \sim \lambda^{-\mu_\parallel}
\end{equation}
is the correlation time in the stationary state for $p=1/2$ and
$0<\lambda\ll 1$. Similarly, the corresponding spatial correlation length
is expected to scale as
\begin{equation}
\xi_\perp \sim \lambda^{-\mu_\perp}\,,
\end{equation}
where $\mu_\perp=\mu_\parallel/z$. To verify this conjecture we measured 
the survival probability $P_{\rm s}(t)$ in seed simulations at $p=1/2$
for two different values of $\lambda$. According to standard scaling theory this quantity
should obey the scaling form
\begin{figure}
\includegraphics[width=80mm]{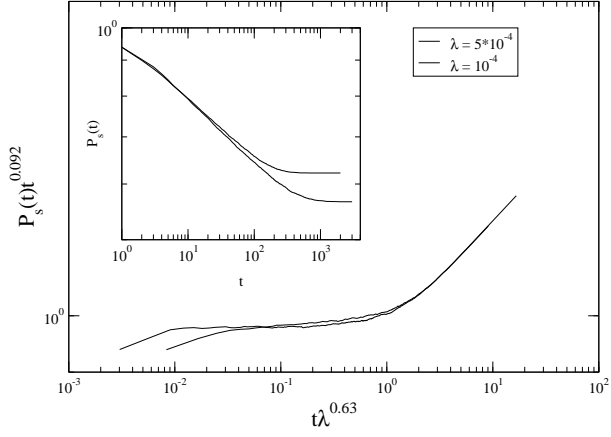}
\caption{
\label{figsurvcollapse}
Data collapse of the survival probability.
}
\end{figure}
\begin{figure}
\includegraphics[width=80mm]{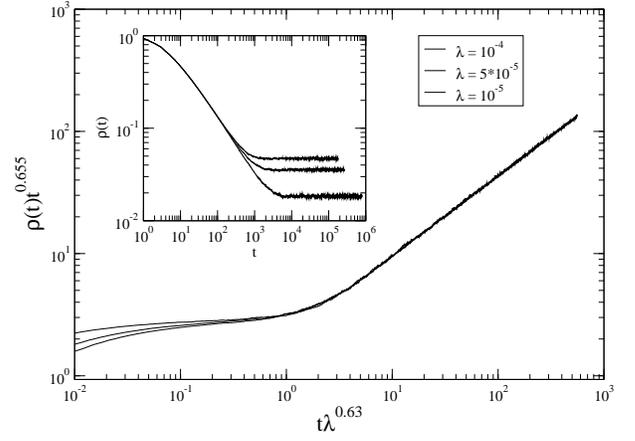} \vglue 4mm
\includegraphics[width=80mm]{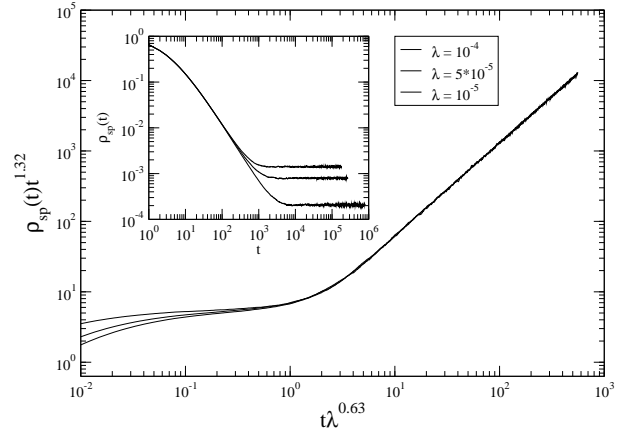}
\caption{
\label{figcollapse}
Data collapse of $\rho$ and $\rho_{\rm sp}$ based on the scaling forms
(\ref{scalingform1}),(\ref{scalingform2}) using the exponents of
Eqs.~(\ref{alphas}),(\ref{mu_numerical}).
}
\end{figure}
\begin{equation}\label{P_s_lambda}
P_{\rm s}(t) \simeq t^{-\delta} \, \Psi(\lambda t^{1/\mu_\parallel})\,,
\end{equation}
where $\delta\simeq0.092$ is the density decay exponent 
of the GEP. Eq.~(\ref{P_s_lambda}) implies that the survival probability eventually saturates at a
value $P_{\rm s}(\infty)\sim\lambda ^{\delta\mu_\parallel}$. Using this scaling form we collapse the data sets 
in Fig.~\ref{figsurvcollapse}, obtaining the estimate
\begin{equation}\label{mu_numerical}
\mu_\parallel = 0.63(3)\,.
\end{equation}
Turning to full-lattice simulations  we can 
combine this scaling law with the results of the previous subsection,
leading to the conjecture that $\rho(t)$ and $\rho_{\rm sp}(t)$
obey the scaling forms
\begin{eqnarray}
\label{scalingform1}
\rho (t)&=&t^{-\alpha}\Omega (\lambda t^{1/\mu_\parallel})\\
\label{scalingform2}
\rho _{\rm sp}(t)&=&t^{-\tilde{\alpha}}\tilde{\Omega} (\lambda t^{1/\mu_\parallel})\,,
\end{eqnarray}
where $\Omega(\zeta)$ and ${\tilde{\Omega}}(\zeta)$ are scaling functions with
the asymptotic behavior
\begin{equation}
\Omega(\zeta) =
\begin{cases}
const & \text{ for } \zeta\to 0 \\
\zeta^{\alpha\mu_\parallel} & \text{ for } \zeta\to\infty
\end{cases}
\end{equation}
Again this implies that eventually the densities reach stationary values
$\rho(\infty)\sim\lambda ^{\alpha\mu_\parallel}$ and $\rho_{\rm
  sp}(\infty)\sim\lambda ^{\tilde{\alpha}\mu_\parallel}$.  
Although there are deviations for small $t$ these scaling forms lead to
reasonable data collapses, as shown in Fig.~\ref{figcollapse}.

We now suggest an explanation for the numerically determined value of
$\mu_\parallel = 0.63$. Initially the process behaves as a critical GEP until
mutations become relevant at a typical time $\xi_\parallel$. Our argument is
based on the assumption that $\xi_\parallel$ scales in the same way as the
typical time at which the first mutation occurs. With the mutation probability $\lambda$ one needs on
average $\lambda ^{-1}$ infections until the first mutation occurs. As the
process initially behaves as a critical GEP the number of infections grows as
$\int dtN(t)\sim t^{\theta +1}$. Hence we are led to
\begin{equation}
\xi_\parallel\sim\lambda^{-\frac{1}{\theta +1}}\,,
\end{equation}   
with $\theta=d/z-2\delta -1=0.587$. This implies the scaling relation
\begin{equation}
\mu_\parallel = \frac{1}{\theta +1}\simeq\frac{1}{1.587} = 0.630
\end{equation}
which is in perfect agreement with the numerical estimation in Eq.~(\ref{mu_numerical}).
%
\subsection{Curvature of transition line at the GEP point}
\label{seccurvature}
So far the numerical analysis suggests that in the vicinity
of the GEP point the epidemic process with mutations is
invariant under scaling transformations of the form
\begin{eqnarray}
x &\to& x' = b \, x \\
t &\to& t' = b^z \, t \\
\label{pscaling}
{\Delta_{\rm p}} &\to& {\Delta_{\rm p}}' = b^{-1/\nu_\perp} \,{\Delta_{\rm p}} \\
\label{lambdascaling}
\lambda &\to& \lambda' = b^{-1/\mu_\perp} \, \lambda \,,
\end{eqnarray}
where $b$ is a scaling factor, ${\Delta_{\rm p}}=p-1/2$, $\delta$ and
$z$ are the critical exponents of GEP, 
and $\mu_\perp=\mu_\parallel/z$.
In addition, the order parameters have to be rescaled appropriately.
In seed simulations this means that
\begin{equation}
P_{\rm s}(t) \to {P'}_{\rm s}(t') = b^{-\delta z} P_{\rm s}(b^z  t) \,,
\end{equation}
leading to the combined scaling form
\begin{equation}
P_{\rm s}(t,\Delta_{\rm p},\lambda)=t^{-\delta}\Phi(\Delta_{\rm p} t^{1/\nu_\parallel},
\lambda t^{1/\mu_\parallel} )
\end{equation}
in the vicinity of the GEP point.
Similarly, in full-lattice simulations one would have to rescale
\begin{eqnarray}
\rho(t) &\to& \rho'(t') = b^{-\alpha z} \rho(b^z t) \\
\rho_{\rm sp}(t) &\to& {\rho'}_{\rm sp}(t') =
 b^{-\tilde{\alpha} z} \rho_{\rm sp}(b^z t) 
\end{eqnarray}
provided that the conjecture of asymptotic power-law behavior
in Eq.~(\ref{powerlawconjecture}) is correct. This would then lead to the
scaling forms
\begin{eqnarray}
\rho(t,\Delta_{\rm p},\lambda) &=& 
t^{-\alpha}\Omega(\Delta_{\rm p} t^{1/\nu_\parallel},
\lambda t^{1/\mu_\parallel} )\\
\rho_{\rm sp}(t,\Delta_{\rm p},\lambda) &=& 
t^{-\tilde{\alpha}}\tilde{\Omega}(\Delta_{\rm p} t^{1/\nu_\parallel},
\lambda t^{1/\mu_\parallel} )\,.
\end{eqnarray}
As usual in the theory
of critical phenomena, the phase transition line itself has to be
invariant under scaling transformations. Comparing Eqs.~(\ref{pscaling}) and~(\ref{lambdascaling}) we are
led to the conclusion that the form of the transition line for
small values of $\lambda$ is given by 
\begin{equation}
\label{curvature}
{\Delta_{\rm p}} \;\sim\; \lambda^\gamma \,,
\end{equation}
where
\begin{equation}
\label{prediction}
\gamma=\frac{\mu_\perp}{\nu_\perp} =\frac{\mu_\parallel}{\nu_\parallel}
\simeq \frac{0.63}{1.506} = 0.42\,.
\end{equation}
Since $\frac{\partial {\Delta_{\rm p}}}{\partial \lambda} \sim \lambda^{-0.58}$
the phase transition line indeed terminates 
at the GEP point with an infinite slope. 

\begin{figure}
\includegraphics[width=80mm]{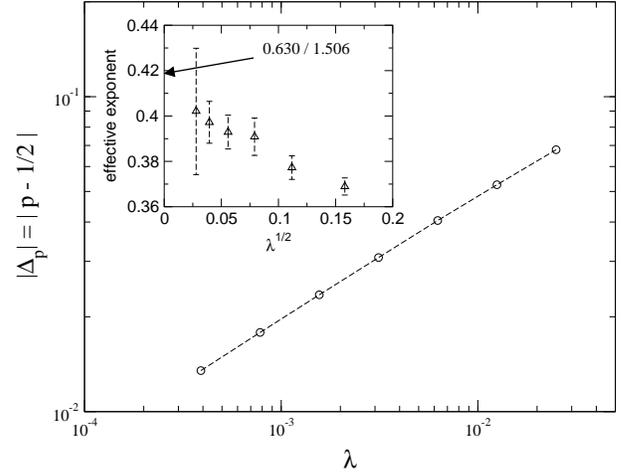}
\caption{
\label{figcurvature}
Double-logarithmic plot of the phase transition line $|\Delta _{\rm p}(\lambda)|$
for $\lambda\ll 1$. The local slope of this curve leads to the effective exponent
which is shown in the inset together with its error bars. The extrapolation of
the effective exponent for $\sqrt{\lambda}\rightarrow 0$ leads to an estimation of
the value of $\gamma$ in Eq.~(\ref{curvature}) which is in agreement with the
prediction of Eq.~(\ref{prediction}).    
}
\end{figure}

In order to confirm the relations~(\ref{curvature}),(\ref{prediction})
we plotted $|{\Delta_{\rm p}}|$ versus $\lambda\ll 1$ in a double-logarithmic
representation in Fig.~\ref{figcurvature}. The local slope of this
curve leads to the effectve exponent (inset in Fig.~\ref{figcurvature}) which can be
extrapolated and leads to the estimation $\gamma \approx 0.41(3)$, in agreement
with the prediction in Eq.~(\ref{prediction}).
%
%
\section{DISCUSSION AND CONCLUSION}
\label{conclusion}
In this paper we have introduced a minimal model for epidemic spreading
with immunization and mutations. Apart from the probabilities for first
infections and reinfections $p$ and $q$
the model is controlled by a probability $\lambda$
that a transmitted pathogen mutates, creating a new pathogen which was
not involved before. The model includes the GEP ($\lambda=0,q=0$) and
DP ($\lambda=1$ or $q=p$) as special cases.

Restricting the analysis to the case of perfect immunization and two
spatial dimensions we have
shown that the transition between survival and extinction in our model
belongs to the universality class of DP everywhere along the phase
transition line except for the point where the model reduces to critical GEP. 
In the vicinity of the GEP point even a small mutation probability drives 
the system away from criticality into a fluctuating active state. The
crossover from GEP to DP can be described in terms of a suitable
scaling theory, which involves a new exponent $\mu_\parallel$.
This crossover exponent also determines the form of the phase transition
line in the vicinity of the GEP point. We suggested an explanation
for the value of $\mu_\parallel$ which turns out to be in perfect agreement
with the numerical analysis. 

Although the model presented here is highly idealized (individuals on
a square lattice, homogeneous infection probabilities, nearest-neighbor 
infections etc.) there is an important conclusion to be drawn
regarding realistic spreading of epidemics in nature. 
As in the model, realistic epidemic spreading starts at 
a certain threshold determined by various parameters such as the average 
susceptibility, the interaction frequency, and the degree of immunization
and/or vaccination. Mutations weaken the effect of immunization, thereby
decreasing this threshold. An important message of our paper is that for a
population which is mainly stabilized by immunization and/or vaccination this
threshold varies {\em nonlinearly} with the mutation rate, in the present case
roughly as the square root of $\lambda$. Thus even a small rate of
mutations can significantly weaken the stability of a population 
at the onset of epidemic spreading.

As a possible extension of the present study it would be interesting to
investigate whether these properties can also
be observed in epidemic processes with long-range infections~\cite{Levy}
if mutations are introduced. Moreover, it would be interesting to study
the surface critical behavior at the system's boundaries~\cite{FHL01}.

\vglue 5mm

\noindent {\bf Acknowledgments:} 
We would like to thank M. L\"assig for bringing our attention to epidemic
spreading with mutations.
The simulations were partly performed on the ALICE parallel computer at the
IAI in Wuppertal. Technical support by B. Orth and G. Arnold 
is gratefully acknowledged.


\end{document}